**The Engineer's Dilemma: A Review of Establishing a Legal Framework for Integrating Machine Learning in Construction by Navigating Precedents and Industry Expectations**

M.Z. Naser, PhD, PE[1,2]
[1]School of Civil and Environmental Engineering & Earth Sciences, Clemson University, USA
[2]Artificial Intelligence Research Institute for Science and Engineering, Clemson University, USA
E-mail: mznaser@clemson.edu, Website: www.mznaser.com

**Abstract**

Despite the widespread interest in machine learning (ML), the engineering industry has not yet fully adopted ML-based methods, which has left engineers and stakeholders uncertain about the legal and regulatory frameworks that govern their decisions. This gap remains unaddressed as an engineer's decision-making process, typically governed by professional ethics and practical guidelines, now intersects with complex algorithmic outputs. To bridge this gap, this paper explores how engineers can navigate legal principles and legislative justifications that support and/or contest the deployment of ML technologies. Drawing on recent precedents and experiences gained from other fields, this paper argues that analogical reasoning can provide a basis for embedding ML within existing engineering codes while maintaining professional accountability and meeting safety requirements. In exploring these issues, the discussion focuses on established liability doctrines, such as negligence and product liability, and highlights how courts have evaluated the use of predictive models. We further analyze how legislative bodies and standard-setting organizations can furnish explicit guidance equivalent to prior endorsements of emergent technologies. This exploration stresses the vitality of understanding the interplay between technical justifications and legal precedents for shaping an informed stance on ML's legitimacy in engineering practice. Finally, our analysis catalyzes a legal framework for integrating ML through which stakeholders can critically assess the responsibilities, liabilities, and benefits inherent in ML-driven engineering solutions.

**1.0 Introduction**

Machine learning (ML) continues to demonstrate substantial promise in various engineering domains. However, professional regulations and industry standards have not addressed ML-based methodologies with the same rigor used for conventional practices, if not at all [1]. This situation prompts questions about the legal and technical justifications for employing ML models when regulatory directives are silent [2]. The same also raises questions with regard to the legality of integrating ML models as engineering codes, which have historically guided design procedures, are now being challenged via algorithmic techniques that introduce new forms of evidence and risk profiles [3].

One prime example of engineering codes can be seen in the use of building codes and testing standards. In most jurisdictions, building codes follow a prescriptive framework that anticipates specific materials and configurations [4,5]. These codes evolve incrementally with scientific advancements, but the incorporation of ML remains unsettled. This gap raises questions about the standard of care: if ML is not codified, engineers risk liability for negligence, yet ignoring proven models might also expose them to technical challenges that may hinder innovation [6].

On the one hand, legislative justification can provide a framework for defending ML usage, wherein engineers can rely on overarching statutes that often employ flexible language regarding duty of care [7]. Such statutes grant engineers latitude to adopt emerging technologies that offer demonstrable improvements in safety. Further, courts have long employed reasonableness standards in negligence claims, and these can also extend to ML, provided engineers validate model outputs and address foreseeable risks [8].

The above infers that legal precedents on novel methods help define how ML might be evaluated. For example, under the Daubert standard in the United States (as well as internationally comparable standards, including Frye and Mohan tests [9,10]), expert testimony is admissible if it is based on sufficient facts, employs reliable principles, and applies these principles reliably to the facts of the case. [11]. Though not a code, Daubert-like frameworks could inform courts in assessing whether ML predictions align with accepted scientific practice. Therefore, engineers can argue that if a model is peer-reviewed, transparent, and generally accepted, it matches the reliability of conventional calculations. Although case law directly on ML in engineering is limited, lawsuits involving algorithmic decisions in other domains suggest how courts may interpret liability [12]. Judges have been noted to scrutinize testing adequacy and the foreseeability of failures in consumer products with autonomous features [13]. Hence, engineers can build on these examples to reduce legal uncertainties.

Away from courts, efforts within standard-setting bodies and professional organizations indicate a growing awareness of the need to formalize ML methodologies in engineering codes [14]. For example, organizations such as the American Society of Civil Engineers (ASCE) and the Institution of Structural Engineers (IStructE) have considered data-driven methods but lack explicit ML provisions [15,16]. A recurring theme is the suggestion that ML tools should undergo rigorous third-party benchmarking, similar to how finite element software is validated via standardized problems or physical tests [17]. Some researchers propose that a repository of accepted datasets and standardized performance metrics would encourage reproducibility and build trust in ML applications [18]. Nonetheless, these proposals have yet to be implemented systematically, which, in a way, reflects the continued tension between innovation and the cautious approach typical of safety-critical disciplines. As engineers rely on professional judgment, drawing analogies to accepted computer-based analyses (e.g., finite element or computational fluid dynamics models) can encourage ML's acceptance [19].

On a more positive note, insights from other regulated domains demonstrate that industry standards can be adaptable. Healthcare regulators, for instance, now recognize certain AI-based diagnostic tools under existing safety frameworks [20]. In addition, a look into precedents from the automotive sector can offer further parallels. Here, autonomous vehicle regulation has progressively formalized requirements for sensor validation, data collection, and real-time decision-making [21]. Although the mechanical complexities differ, the underlying challenge of legally endorsing algorithmic control in high-stakes scenarios still resonates with ML in engineering [22].

Another look into code adaptation can stem from how traditional engineering codes remain static documents mostly (but can be updated at times), whereas ML models can evolve as new data/algorithms are acquired [23]. This dynamic characteristic raises questions about potential divergence from originally validated states. This scenario can build on the fact that courts and legislatures have tackled questions about software updates, liability distribution between manufacturers and operators, and the sufficiency of real-world testing [24]. Thus, engineers can cite these precedents to argue that statutes can accommodate ML without undermining public welfare by integrating continual learning and developing robust validation/verification standards [25]. More realistically, integrating ML can be undertaken via pilot programs or provisional approvals to enable incremental integration where ML tools move from restricted pilot uses to broader acceptance – similar to the approaches used in automotive or pharmaceutical industries [26].

Despite such a positive look, it must be stressed that a direct transference of existing frameworks across domains can be complicated by the unique nature of engineering projects, where each structure or system can differ substantially in materials, design complexity, and local constraints [27]. Here, a broader acceptance could build on the momentum of the recent efforts of various organizations. More specifically, bodies such as the International Organization for Standardization (ISO) and the International Electrotechnical Commission (IEC) have started to address ML risk management [28]. Their guidelines often aim for global applicability and reducing fragmentation across jurisdictions. However, national legislative differences persist, and engineering licensure remains largely jurisdiction-specific [29]. While a unified international code for ML in engineering might be ideal, it may not be likely to be achieved in the near term. Instead, engineers often combine local building codes with overarching ISO frameworks. Extending this strategy to ML would require explicit procedures for validating models under varied conditions.

Implementing ML in engineering also raises data governance and cybersecurity concerns [30]. For instance, unauthorized access to ML systems could compromise design integrity, which may pose serious risks to public safety. These considerations add another layer to legislative justifications, as codes may eventually require encryption standards, secure data storage protocols, or redundancy measures to mitigate cyber threats [31]. The complexity of ML pipelines, involving data acquisition, cleaning, model training, and continuous updates, suggests that future codes might incorporate guidelines on end-to-end lifecycle management to ensure that security and privacy are integral aspects of ML integration [32,33].

This article adopts a position-based perspective on the legal, doctrinal, and legislative justifications for using ML when engineering codes remain silent. We highlight analogical reasoning from other regulated sectors, examine liability issues under tort and product liability principles, and consider emergent ML-specific legislation. By synthesizing these domains, we propose a framework for engineers, policymakers, and legal practitioners to evaluate and incorporate ML responsibly. We aim to ensure that ML adoption in engineering remains legally defensible and conducive to advances in safety and efficiency.

## 2.0 Legislative bodies and standard setting in engineering contexts
This section presents some of the legislative bodies and standard-setting organizations and showcases how they can influence the integration of ML.

Historically, building codes and technical standards emerged as a response to societal needs for safe, reliable structures [34]. In many jurisdictions, compliance with widely recognized standards like the ASCE 7 or the Eurocodes is important, as departures from these references can invite legal scrutiny, especially if an incident occurs. These standards have historically emphasized validated mathematical models, prescriptive design formulas, and well-tested materials [35]. Thus, the potential introduction of ML algorithms confronts conventional model demonstration and acceptance [36]. Today, the rapid incorporation of ML into engineering workflows presents a new challenge for these entities as they must determine how to integrate ML into existing frameworks that were largely developed for design procedures and materials testing – some of which may fall under empirical approaches, physics-based approaches, etc.

As these bodies typically operate through committees composed of academic experts, industry professionals, and sometimes government representatives, legislators rely on subject-matter experts to provide guidance on ML-based methods. In response, committees must now decide whether to expand existing documents or draft new standards that address data collection, model training, performance metrics, and lifecycle for ML-driven solutions. This deliberation often requires balancing innovation against the risk of endorsing methods that may lack long-term track records. Therefore, without a clear consensus from professional bodies, lawmakers may be reluctant to endorse or mandate ML usage [37]. Consequently, engineering firms seeking legal certainty might forgo ML or apply it only in research or non-critical tasks, limiting the technology's broader adoption [38]. However, with time and continued advancements in ML, legislative bodies are likely to face pressure from industry stakeholders and insurance underwriters to clarify liability boundaries [39]. This uncertainty can result in higher premiums or coverage exclusions that deter engineering firms from experimenting with ML. Thus, legislators can reduce ambiguity by establishing transparent guidelines, endorsing standard-setting organization protocols, or perhaps opting to avoid using ML.

Here, it would be of interest to note how some legislative bodies have introduced performance-based codes that offer more flexibility for novel methods (see IBC Section 104.11, 2018 edition [4]). In principle, performance-based frameworks lower barriers to innovation because they do not mandate specific calculations or materials and focus instead on achieved outcomes. Such provisions may permit any approach that demonstrates equivalent or superior safety compared to prescriptive norms [40]. So, in the absence of explicit ML provisions, engineers may leverage performance-based clauses to justify employing ML models. Engineers can leverage ML in designs if they substantiate the results through peer-reviewed studies, simulations, or experimental data. This can be tackled by systematically documenting assumptions, validation procedures, sensitivity analyses, etc. This framework parallels fire engineering, where performance-based codes allow computational fluid dynamics simulations of fire spread, provided they meet established accuracy benchmarks [41]. The challenge for standard-setting bodies is determining how to define equivalency or superiority for ML-driven methods. For instance, a performance-based code might require a reliability index above a certain threshold, which leaves the question of how to verify an ML model's statistical consistency or performance on other design aspects as well.

Another worthy discussion could arise by examining how sector-specific legislation may influence the integration of ML (say, in highly regulated industries such as the Federal Aviation Administration (FAA)). Such agencies often impose rigorous certification protocols for new technologies [42]. Although these bodies may not directly oversee specific engineering practices (such as civil and construction engineering), their acceptance criteria for advanced software can set a precedent for other domains. Suppose an ML-based diagnostic tool obtains FAA approval for designing aircraft components [43]. In that case, civil engineering standard-setters may view that certification as evidence of reliability and could entertain a parallel approach for structural applications[1].

In many regions, engineers are licensed by a professional board that enforces codes of ethics and technical standards. Licensure statutes typically demand engineers adhere to recognized procedures and maintain competence in emerging areas of their field. While professional licensure boards are often thought of passively with regard to legislation, they can also indirectly shape standard-setting by enforcing codes of ethics and continuing education requirements [44]. For example, in jurisdictions where professional engineers must demonstrate ongoing competence, licensure boards can require or recommend training on ML methods, including the pitfalls of ML. While licensure boards seldom draft technical standards, they can accelerate or impede ML adoption by endorsing certain best practices as part of continuing education curricula [45]. If these boards unify around a set of guidelines, legislative bodies may adopt them into official code language and, by extension, formalize ML usage in engineering.

At the moment, and in the absence of explicit mandates, soft law instruments such as consensus reports, manuals of practice, and technical guides could gain importance [46]. For example, committees, task forces, and expert working groups within professional bodies may publish white papers outlining recommended ML model development and validation procedures. While these documents lack the force of statutory law, they can still provide some influence in legal disputes. Courts may treat them as indicative of industry best practices, and legislative bodies often look to these technical guides when formulating or revising statutes [47]. Over time, an accumulation of well-regarded soft law documents can motivate official standard-setting organizations to officialize ML-based standards [27]. Table 1 Summarizes the main concepts covered herein.

Table 1 Main concepts covered in this section

| Category | Description | Impact on ML Integration |
|---|---|---|
| Building Codes & Standards | Building codes (e.g., ASCE 7, Eurocodes) provide safety and reliability frameworks but lack explicit ML guidelines. | ML struggles with acceptance due to the traditional reliance on validated mathematical models and prescriptive formulas. |
| Committee Decision-Making | Committees composed of experts debate integrating ML into standards and aim to balance innovation and risk. | Decisions on ML guidelines depend on achieving consensus among committee members. |
| Legislative & Regulatory Challenges | Lack of clear regulations deters ML adoption; insurance and liability concerns affect its use in critical applications. | Firms may hesitate to use ML due to legal uncertainty, which can adversely affect broader adoption. |
| Performance-Based Codes | Performance-based codes allow novel approaches if they meet equivalent or superior safety standards. | ML may be justified under performance-based frameworks if it demonstrates reliability and accuracy through validation. |
| Sector-Specific Regulations | Some agencies may enforce strict certification for new tech. | ML tools approved in regulated industries may serve as a precedent for civil engineering applications. |
| Professional Licensure Boards | Licensure boards enforce ethics and continuing education, potentially shaping ML adoption via training mandates. | Mandated ML training for engineers could accelerate industry-wide adoption and regulatory acceptance. |
| Soft Law Instruments | Consensus reports, white papers, and expert guidelines, while not legally binding, may influence regulations over time. | Soft law instruments may serve as interim standards before official regulations catch up with ML developments. |

---

[1] It is also equally possible that skepticism can arise if these agencies express concerns about model performance.

## 3.0 Legal precedents and legislative justifications

Although traditional engineering codes (such as the International Building Code (IBC), etc.) do not expressly mention algorithmic decision-making, the core principles of safety, reliability, and professional responsibility remain expected [48]. In light of this development, engineers and regulators must identify credible legal and legislative justifications to support using ML. The legal framework surrounding engineering practice rests on foundational doctrines of liability, professional responsibility, and statutory compliance [49]. As courts cope with this emergent technology, certain precedents and legislative principles can guide the engineering community in understanding their responsibilities and potential liabilities. This section hopes to address some of those principles – also see Fig. 1.

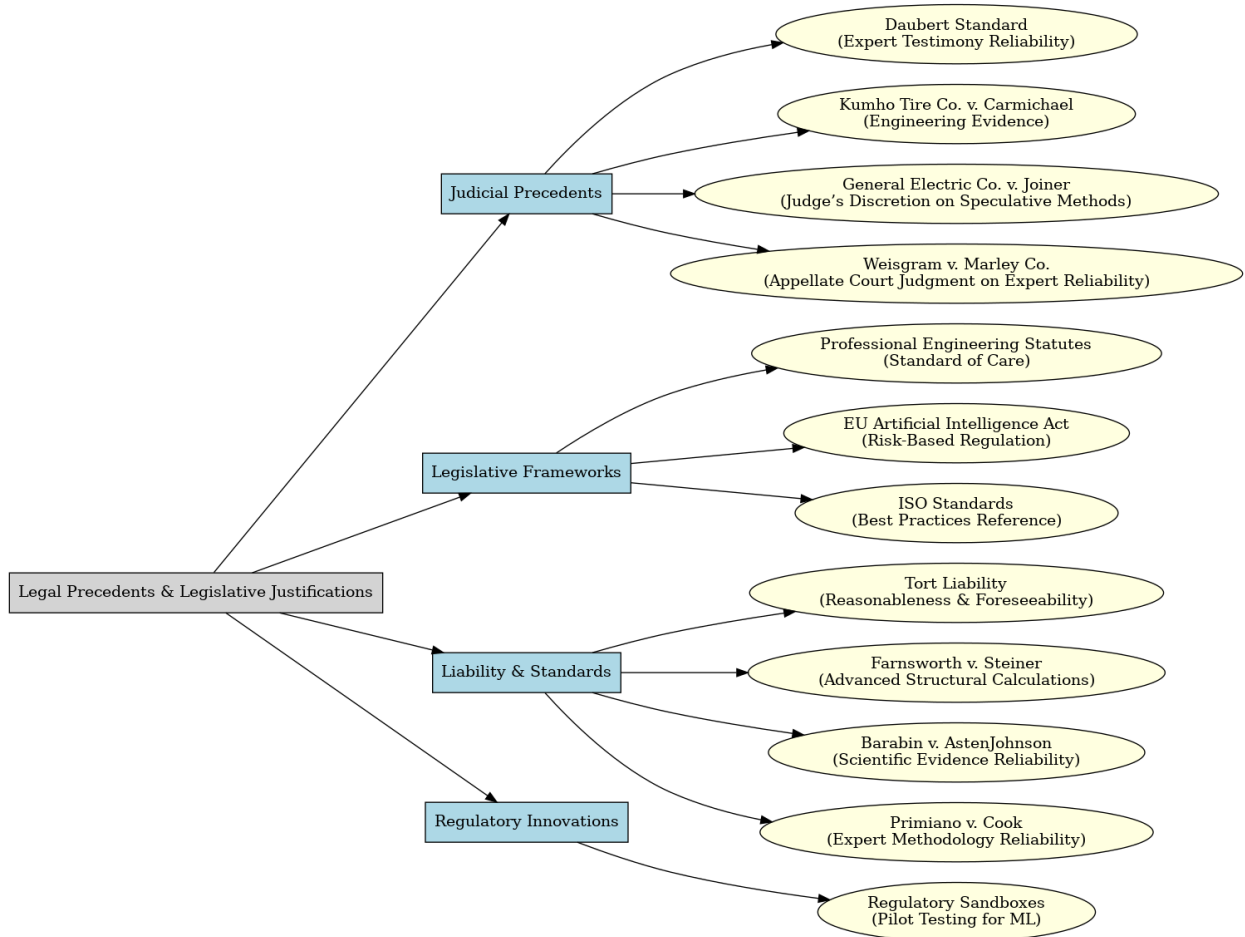

Fig. 1 Legal precedent and legislative justifications for ML in engineering

At a basic level, courts have begun to confront scenarios where computational models serve as the basis for expert testimony [50]. This emerging body of case law provides insight into how ML could be justified in engineering contexts even without explicit code provisions. To start, and reflecting on a notion mentioned in the introduction section, the Daubert standard for expert witness testimony highlights how courts might evaluate the reliability of ML-based engineering decisions. In Daubert v. Merrell Dow Pharmaceuticals, Inc. (509 U.S. 579), the Supreme Court outlined criteria for judging expert testimony's scientific validity and relevance by emphasizing peer review, testability, and acceptance in the relevant scientific community [51]. Although this case concerned pharmaceutical evidence, its rationale was extended to engineering experts in Kumho Tire Co. v. Carmichael (526 U.S. 137 [52]). Another closely aligned case is General Electric Co. v. Joiner (522 U.S. 136 [53]), where the Supreme Court reaffirmed that trial judges have the discretion to exclude expert evidence grounded in unsupported or overly speculative methodologies and required a clear scientific connection between data and conclusions. Another significant ruling is Weisgram v. Marley Co. (528 U.S. 440 [54]), which clarified that appellate courts may enter judgment if expert testimony fails to meet Daubert's reliability requirements. These precedents indicate that courts could potentially assess whether ML-generated methods follow recognized scientific or engineering methodologies. This suggests that ML, when subjected to the same scrutiny as traditional computational methods, may be viewed as an admissible and reliable basis for professional engineering decisions [55].

Legislative justifications often hinge on the public welfare and safety objectives preserved in existing statutes that govern engineering practice. In the United States, for instance, most states have enacted professional engineering statutes that emphasize a *standard of care* that revolves around the reasonably prudent engineer concept (see [56]). These laws do not prescribe an engineer's specific tools or models; rather, they require that professional judgments be made with high competence and caution. Thus, ML models can fit within this framework if engineers demonstrate that their use of ML constitutes an acceptable professional practice, as demonstrated in peer-reviewed studies, or showcase that ML-driven optimization has undergone the same or greater level of verification as FE analyses [57].

From a comparative perspective, European legal systems offer additional guidance on how ML might be integrated into engineering practices. The proposed European Union Artificial Intelligence Act categorizes ML systems based on risk levels [58]. Though primarily aimed at regulating consumer-facing AI, the legislative outlines a broader risk-based approach that can extend to infrastructure settings. Under this framework, an ML model used in high-stakes projects might fall under a high-risk category and, hence, is likely to trigger requirements concerning data governance, record-keeping, and human oversight. While not specifically targeting building codes, the Act's emphasis on transparency and accountability could encourage engineering bodies to adopt standardized validation procedures and detailed documentation of their ML analyses. A similar observation can also be seen in transnational standards (i.e., ISO), which provide a foundational overview of ML. While such standards do not carry the force of law, they could serve as references for regulatory agencies when determining whether engineers have adhered to recognized best practices.

*Tort liability* cases offer a precedent for applying advanced analytics in engineering. One might have drawn an analogy to adopting FE methods several decades ago: once considered cutting-edge, these computational techniques are now broadly accepted, and courts treat them as standard engineering practice. Courts consistently evaluate whether the defendant (in this case, the engineer) acted reasonably under the circumstances, looking at whether the professional conformed to the standard practices of the field and exercised competence [59]. In such instances, courts typically examine engineers' processes, whether known risks were disclosed, and whether there were mechanisms to revert to manual control or conventional methods. Suppose an ML-driven design program underestimates load variability, which can lead to structural deficiencies. In that case, a court will inquire whether the engineer should have foreseen that the model might be prone to certain data limitations.

Another example can be seen in a structure's design that might incorporate a hybrid approach, where ML-driven optimizations are reviewed through a conventional code-based lens to create redundancy that reduces litigation exposure. Lawmakers may view such hybrid strategies as prudent to balance technological progress with fail-safes aimed at public welfare. A more technical example can be seen in Farnsworth v. Steiner (601 P.2d 266, Alaska [60]), where advanced structural calculations were deemed admissible as expert evidence. In Barabin v. AstenJohnson, Inc. (700 F.3d 428, 9th Cir. [61]), the court reversed a judgment because expert testimony on scientific matters was not properly vetted under reliability standards, which reflects the heightened probe courts apply to technical evidence. Similarly, Primiano v. Cook (598 F.3d 558, 9th Cir. [62]) demonstrated that expert conclusions must stem from reliable methodologies in the relevant professional community.

Based on these precedents, if an ML model is used for predictive maintenance, and an anomaly is overlooked because the model was insufficiently trained on outlier data, an injured party could argue that the system failed to foresee a reasonably probable event [63]. Engineers can counter by demonstrating that they adopted recognized data enrichment or robust training strategies, thereby satisfying the foreseeability requirement. Conversely, if the event was genuinely unforeseeable or caused by an unprecedented confluence of factors, engineers might argue that no reasonable professional could have anticipated it, thereby reducing or negating liability. Moreover, tort doctrines emphasize foreseeability and causation [64]. Courts also analyze causation by determining whether the ML error precipitated the damage. The above patterns illustrate the necessity of robust documentation and fallback systems when employing ML in engineering [65].

A more recent area of relevance is the use of *regulatory sandboxes*, which some state legislatures have adopted in other industries (particularly in finance) to facilitate innovation in a supervised environment, as seen in Arizona's first fintech sandbox [66]. While not directly applicable to building codes, the sandbox concept allows regulated entities to test new technologies under agency oversight with limited liability or alternative compliance structures [67]. This may serve as a key precedent wherein engineering boards or governmental bodies create similar programs tailored to ML-driven engineering methodologies. In this case, judicial systems often treat such recognized pilot programs as evidence that a technology complies with existing standards [68].

### 4.0 Negligence, liability doctrines, and accountability

In engineering, liability doctrines have traditionally centered on *negligence*, *liability*, and *accountability*. These established frameworks face novel pressures as ML becomes integrated into engineering workflows. Here, we shed light on such pressures – see Fig. 2.

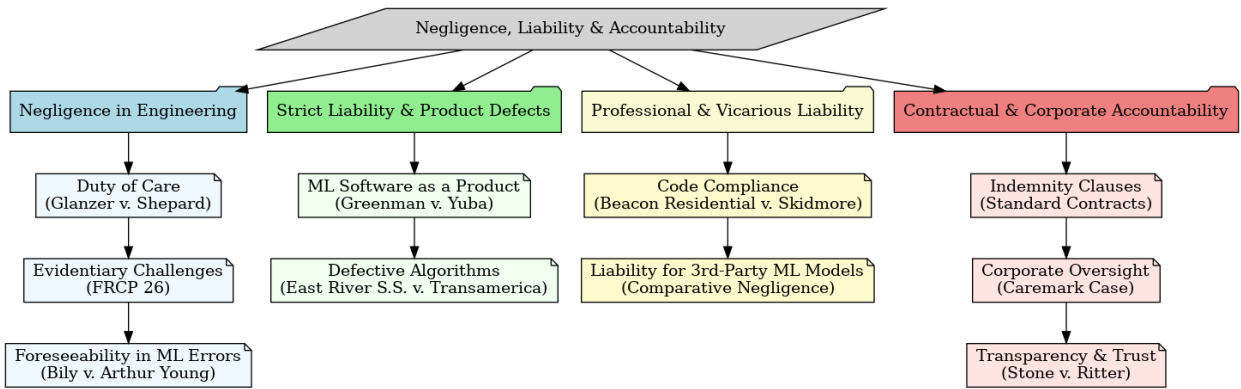

Fig. 2 Flowchart of possible negligence, liability doctrines, and accountability with regard to ML

*Negligence* in professional engineering derives from the broader tort principle that a defendant is liable for harm caused by a breach of a duty of care [69]. The professional standard of care is typically grounded in what a reasonably prudent practitioner in the same field would do under similar circumstances (see Glanzer v. Shepard [70]). This translates to expecting the engineer to perform services following accepted engineering principles and practices. Determining whether an engineer has breached this standard can be more challenging where ML is involved, particularly if the engineer relies on proprietary or blackbox algorithms. Similarly, the causal chain might be less transparent when ML systems are involved, especially when the model operates via deep neural networks that produce intermediate variables and weighting factors not readily interpretable by human reviewers. Thus, a question pertaining to the evidentiary challenges associated with establishing causation arises [71]. Addressing such a question is possible as discovery rules, including Federal Rule of Civil Procedure 26 in U.S. courts, permit broad requests for source code, validation studies, and internal documentation and may also allow expert witnesses to scrutinize the modeling process to determine where assumptions, training data, or parameter selections might have introduced errors [72].

Another key consideration is *foreseeability*, wherein negligence typically requires that the resulting harm be a reasonably foreseeable consequence of the engineer's conduct. This consideration could be problematic when ML models are trained on historical data that fail to account for unusual or rare events. Two cases shed light on how courts approach foreseeability standards seen in Bily v. Arthur Young & Co. (834 P.2d 745, Cal. [73]), where the California Supreme Court reiterated that expert liability hinges on whether the defendant's conduct meets the benchmark of a reasonably prudent professional emphasized the need for rigor in methodology and disclosure. Likewise, Ultramares Corp. v. Touche (255 N.Y. 170 [74]) clarified that professionals might be liable to third parties if the harm is foreseeable and the professional owed a duty of care. These rulings imply that ML-driven engineering analyses must still align with well-recognized professional standards of diligence and transparency.

An engineer's potential liability under the negligence doctrine also extends to scenarios where the ML model performed as designed but generated outcomes that deviate from standard code-compliant procedures [75]. Even if building codes do not explicitly mention ML, code provisions typically reflect baseline safety and performance requirements. For example, say, an ML model produces a structural detail that fails to comply with a mandatory minimum, such as a specified safety factor for load-bearing columns. In this case, the engineer cannot avoid liability solely by contending that the model's logic indicated an alternative design. Courts rely on code provisions and well-established principles of professional liability and will likely determine that the engineer breached the duty of care by approving a design that did not meet explicit normative standards [76]. A relevant illustration is Beacon Residential Community Assn. v. Skidmore, Owings & Merrill LLP (59 Cal.4th 568 [77]), where design professionals were held accountable when alleged code-related deficiencies contributed to unsafe conditions. The court found that architects and engineers cannot hide behind contractual roles if the project design materially deviates from accepted building standards. In a similar vein, Bilt-Rite Contractors, Inc. v. The Architectural Studio (866 A.2d 270, Pa. [78]) confirmed that design professionals owe a duty of care to downstream users for code and safety compliance. These precedents underscore that ML outputs must meet unambiguous regulatory thresholds to avoid liability.

While ordinary negligence revolves around the engineer's failure to meet the standard of care, *strict liability* traditionally applies in product defect cases or inherently dangerous activities regardless of the producer's intent or negligence. In engineering, strict liability typically applies to physical products such as building materials, structural components, or machinery if they pose undue hazards to users or the public [79]. Thus, this line between a service (e.g., a structural design calculation) and a product (e.g., a ML tool) sold to multiple third parties can become blurry. For example, a software developer who creates an ML-based structural analysis program might, in theory, face *product liability* claims if the algorithm's inherent defect leads to unsafe designs. In such an example, if a building's structural design relies on a proprietary ML tool that later proves defective, claimants may seek recourse against both the engineering firm and the software vendor. In this case, courts look for evidence that the software was tested and that warnings about limitations were adequately communicated. In this environment, accountability extends to all parties within the supply chain, including external consultants providing ML-based solutions.

As one can see, *product liability* doctrines can further overlap with contract law and warranty claims, particularly when ML models are sold or licensed to engineering firms under specific performance guarantees. Suppose vendors represent their ML tool as "highly accurate" or "industry-leading," but fail to disclose known limitations or error margins. In that case, an injured party may argue that the warranty was breached when the software miscalculates load distributions or fails to detect critical anomalies.

Under strict liability regimes, manufacturers or sellers of a defective product can be held liable for injuries caused by that product, irrespective of negligence. By contrast, an engineering firm that customizes an ML model for its own projects would more likely be subject to professional *malpractice claims*, not strict liability, because it primarily renders a service. Courts have generally recognized this distinction in cases involving professional services by indicating that controlling doctrine remains negligence rather than strict liability, absent an explicit product-based transaction. This can be seen in the case of East River S.S. Corp. v. Transamerica Delaval, Inc. [80], which distinguishes product defects from professional services. Another product liability case is Greenman v. Yuba Power Products, Inc. (59 Cal.2d 57 [81]), where the California Supreme Court established that a manufacturer is strictly liable for injuries caused by defects in its product. The ruling shows that proving negligence is unnecessary if the product is shown to be defective and the defect directly causes harm. One should still note that integrating ML-based systems into similar products can complicate matters by potentially transferring part of the liability onto software developers or data providers, especially if an algorithm's output leads to design flaws that pose safety risks.

*Professional liability* or *malpractice doctrines* add another layer of accountability by focusing on whether the engineering practitioner has upheld the recognized standard of care. This ethical duty is protected in codes such as the National Society of Professional Engineers (NSPE) Code of Ethics [82] and persists regardless of the introduction of automated methods. Simply, if an algorithm's recommendations cannot be explained in court or peer review, practitioners may struggle to demonstrate that they practiced due diligence. By extension, this accountability also extends beyond initial deployment to maintenance and updates: ML models evolve if retrained with new data, implying that ongoing oversight is essential to meet professional standards.

*Vicarious liability* doctrines could also arise when engineers collaborate with sub-consultants, third-party software developers, or data scientists. If the engineer oversees the overall design but relies on an external ML service for structural analysis, questions emerge about who bears responsibility for errors embedded in the outsourced solution [83]. Traditional agency principles may attribute liability to the primary engineering firm if it has supervisory authority over the subcontractor's work. On the other hand, the subcontractor could share liability if it failed to meet its professional obligations or misrepresented the solution's reliability [84]. For instance, if the software developer's negligence in coding or training the model substantially contributed to the defect, and the engineering firm's lack of adequate verification also played a role, courts might approve fault under comparative negligence principles. Thus, clear contracts delineating responsibilities and performance obligations become vital in fairly distributing accountability [85]. Such complexity does not eliminate liability; instead, it necessitates a more granular analysis of causation and fault allocation.

The above may raise a question with regard to *contractual liability* in ML-assisted projects (which can overlap with tort-based theories). While existing standard form contracts in construction might not specifically address ML usage, they invariably impose performance obligations that reflect the professional standard of care. Some engineering contracts may include indemnity clauses or limitations of liability, yet courts typically refuse to enforce contract terms that completely absolve a professional from responsibility for gross negligence or willful misconduct [86]. This tension between contract disclaimers and tort duties highlights how an engineer's accountability cannot be evaded merely by referencing the technology's experimental status.

In addition to legal doctrines, *accountability* intersects with evolving professional ethics and corporate governance norms. As seen earlier, engineering and insurance firms could implement internal policies that mandate peer review or third-party audits of ML models. But, accountability can also stretch past the individuals who directly interact with the ML algorithm. For example, oversight committees may review the adoption of ML methodologies in key projects in large engineering firms to avoid corporate negligence claims. Courts have, in certain contexts, recognized a corporate duty to ensure that employees follow established protocols (see In re Caremark International, Inc. - Derivative Litigation, Del. Chancery [87]). Further, Stone v. Ritter (911 A.2d 362, Del. [88]) refined the standards set in Caremark by clarifying that directors can be personally liable if they fail to implement any reporting or oversight mechanisms or monitor those systems once in place. Transposed to ML setting, a failure to implement adequate procedures for algorithmic validation or a disregard for known biases in training data may implicate the engineer even when disclaimers, without demonstrable oversight, are utilized as these are likely to be deemed too broad or vague [89]. An adjacent dimension is referred to as the *social accountability* of engineering organizations [90]. In this view, public trust in infrastructure safety may decline if widely reported algorithmic failures go unaddressed or lead to catastrophic outcomes.

Consequently, policymakers and professional bodies encourage transparency about the capabilities and limitations of ML systems.

**5.0 Possible legal framework for integrating machine learning into engineering practice**
Developing a structured legal framework is essential for ensuring that ML deployments align with established engineering expectations and norms. Such a framework would address how codes, statutes, and professional regulations can be adapted or extended to incorporate automated decision-making while preserving public trust and minimizing liability risks. This section presents a possible framework that requires careful balancing of technical rigor, transparent governance, and flexibility to accommodate ongoing ML advances – see Fig. 3.

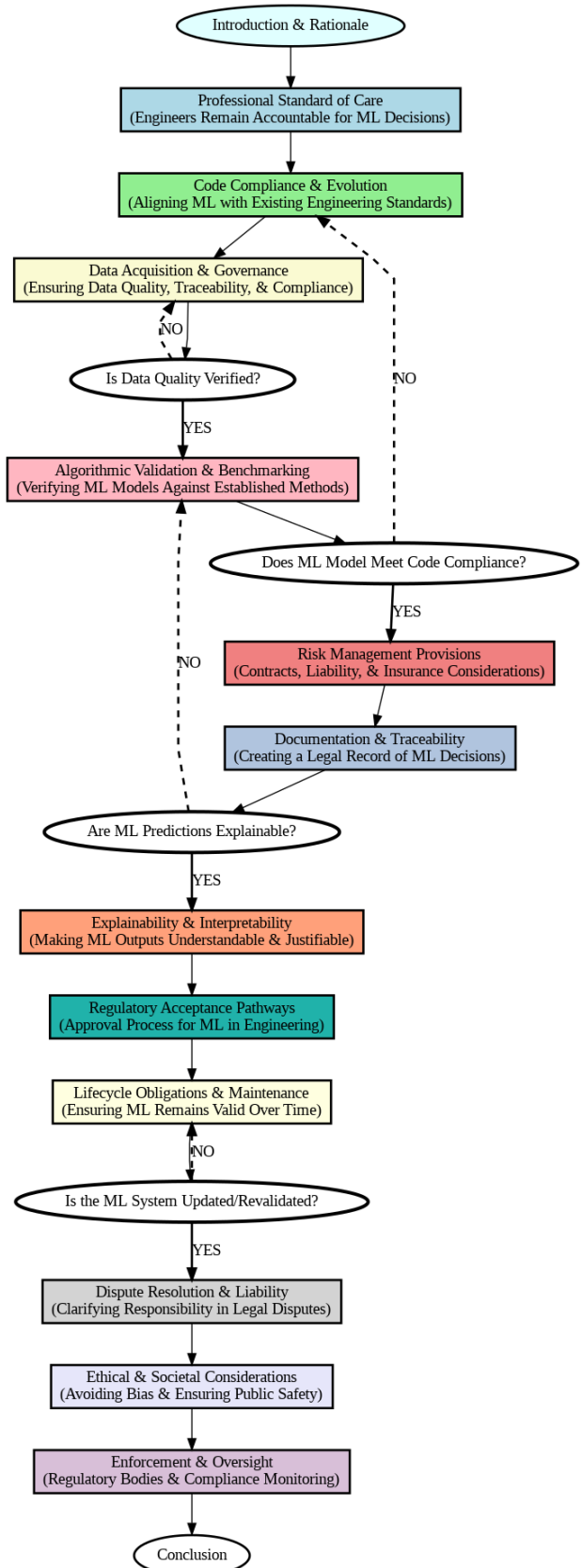

Fig. 3 Flowchart of the proposed framework

As seen in the above review, the professional standard of care is a cornerstone of engineering liability. The legal framework for ML should clarify how traditional concepts of due diligence extend to algorithmic decisions [91]. For instance, design computations relying on proprietary blackbox ML models must include independent validation/common sense steps similar to standard simulation checks in FE methods [92]. Further, the engineer retains ultimate responsibility for verifying that data inputs, model selection, and interpretative outputs do not deviate from established safety margins. Thus, any legal text or model code supplement would stipulate that an engineer's reliance on ML is conditionally permissible only if safeguards are in place [93]. This ensures that an algorithm's opacity does not erode the professional's accountability for the final deliverables.

The above review also notes that while most engineering codes lack explicit language addressing ML, they are flexible enough to accommodate new methodologies if they meet the baseline thresholds via the performance-based approach [94]. Thus, a legal framework should offer guidelines on how ML-based designs can be deemed code-compliant. One potential solution is to create "appendices," "commentaries," or "alternative methodology" sections within existing standards to outline steps for equivalence of performance for ML outputs, possibly through structured test suites or standardized benchmarks that incorporate known structural configurations, boundary conditions, and material properties. To assess consistency, the engineer would compare ML outputs to conventional simulations or real-world measurements [95]. Any significant discrepancies require explanation or model tuning. Guidelines might also incorporate reliability metrics such as confidence intervals or partial safety factors to account for ML uncertainty. A possible approach might entail submitting a preliminary notice (where engineers inform regulatory bodies of their intent to employ ML-based methodologies, identifying the scope and risk classification), followed by a validation dossier (a structured compilation of evidence, i.e., benchmark comparisons, test results, model documentation, etc.). Then, an independent peer review with subject-matter experts confirms that the ML approach meets the required safety objectives [96]. The outcome can be disapproval, approval, or conditional approval. Over time, successful projects might prompt official code committees to write specialized chapters on ML usage, thus gradually shifting ML from an "alternative method" to an acknowledged mainstream approach.

From a legal standpoint, improper data curation can expose engineers and organizations to professional liability claims [97]. Thus, any proposed framework should mandate vigorous data governance policies, including traceability of data sources, metadata verification (e.g., calibration logs), and documented data cleaning protocols. In addition, compliance with privacy statutes must be codified to confirm that data usage does not infringe on privacy or intellectual property rights, which may impose risk-related cases [98]. Furthermore, risk management extends beyond verifying ML's model data or a design's technical correctness. It encompasses contractual stipulations, insurance coverage, and contingency planning for unforeseen model failures. A practical legal framework would specify that contracts define the roles and obligations of engineering firms, software vendors, data providers, and any subcontractors responsible for ML deployment [99]. Further, the framework should encourage engineers to maintain professional liability insurance that explicitly addresses algorithmic errors [100].

The afore-discussed precedents in Sections 3 and 4 note how courts and regulatory bodies frequently require thorough documentation to establish causation and fault [101]. Hence, engineers might face unique evidentiary demands when ML tools are used to show how specific model outputs inform a design choice. Therefore, the legal framework should stipulate that every ML-based decision path is traceable through logs or version control systems. This includes data logs (e.g., timestamps of training sets, data preprocessing steps), model configurations (e.g., hyperparameters, architecture details), validation reports (e.g., performance metrics, sensitivity analyses), model output (e.g., predicted load distribution or stress profiles), etc. Similarly, an auditable record must show changes to data or algorithmic parameters and their rationale if updates occur after initial deployment [102]. Doing so ensures that if disputes arise, an expert witness can reconstruct the engineering logic and verify it against recognized standards of care.

For legal acceptance, engineers must provide domain-based explainability mechanisms [103]. An organic and recommended approach involves "physics-guided ML," wherein domain constraints, such as equilibrium equations, are embedded into the training process to ensure physically viable outcomes [104]. This approach can be essential in jurisdictions where judges or juries demand comprehensible evidence that a design process aligns with classical engineering logic. The legal framework needs to emphasize that engineers should select or develop ML architectures that yield interpretable intermediate results [105].

Since ML models can be continuously retrained with new data, then from a legal standpoint, an engineer cannot simply presume that prior regulatory approvals still apply if the model's architecture or training sets change substantially [106]. Therefore, the framework must define trigger points for re-approval or re-certification, such as significant modifications in the underlying model architecture, additional training on newly acquired data that differs materially from the original sources, discovery of a structural deficiency traceable to an algorithmic shortcoming, etc. This approach allows incremental changes to the ML model to remain legally accountable and not compromise public safety [107].

Some ML-driven projects will inevitably experience disputes with varying root causes of a structural issue or the extent of an engineer's reliance on algorithmic recommendations. The legal framework should clarify how liability is apportioned among the involved parties [108]. Provisions might borrow from well-established principles of comparative negligence or joint-and-several liability, with carefully tailored exceptions. Similarly, any legal framework must address societal concerns, such as disparate impact on vulnerable communities. Thus, the framework could require an "equity review" step where engineers analyze the model's predictions for outlier regions or communities. If any issues arise, corrective measures can be implemented [109].

Eventually, a legal framework demands adequate enforcement mechanisms [110]. For example, government agencies, professional licensing boards, and third-party auditors need to have clear authority to review ML-based designs. Licensing boards could institute continuing education requirements on ML management and evolving best practices [111]. Failure to fulfill these requirements could result in disciplinary actions or penalties.

## 6.0 Future research directions and needs

The integration of ML into engineering opens an array of possibilities that transcend mere efficiency gains. Yet its reliable and legally defensible adoption demands focused research across multiple fronts. A pressing need involves the development of open-source standardized benchmarks and testbeds for verifying ML algorithms in various engineering applications [112,113]. This may stem from similar efforts undertaken in software validation and could help serve a new purpose for comparing/validating ML to existing numerical methods (FE and CFD).

Similarly, investigations into licensing models, peer review mechanisms, and open-data initiatives can be crucial to ensure that these collaborative efforts maintain scientific rigor. Such interdisciplinary collaboration could benefit from including legal scholars to provide guidance on how evolving liability doctrines and privacy regulations intersect with engineering safety imperatives [114]. This could also help to shape consistent governance frameworks. This synergy may extend to the design of advanced risk models that factor in structural uncertainties and algorithmic and cyber threats [115]. Their findings can then inform code updates, ensuring that changes are grounded in practical outcomes rather than theoretical possibilities. ML-based solutions will likely follow a phased acceptance in lower-risk applications before progressing to core structural design tasks.

Explaining ML output stands as another high-priority research area, as their opaqueness complicates regulatory approval and legal defenses [116]. Hence, explainable ML approaches that align model outputs with engineering principles could alleviate concerns about unpredictability [117]. Researchers could explore hybrid models that embed physics-based constraints within data-driven architectures to ensure that outputs respect fundamental laws such as equilibrium and compatibility [118]. These methods strengthen trust among engineers and streamline judicial scrutiny by reducing the interpretability gap by clarifying model reasoning (which is crucial in court settings). Legal legitimacy may suffer if a jury or judge cannot grasp how an algorithm arrives at its predictions, regardless of a model's statistical performance.

A third research priority lies in the lifecycle management of ML models [119]. While conventional engineering design often concludes once construction is complete or machinery is commissioned, ML models can evolve if exposed to ongoing data streams. Future work can define protocols for retraining, version control, and rollback of algorithmic updates. This includes investigating continuous integration pipelines where new data triggers periodic model recalibration [120]. Interested researchers could explore fail-safe mechanisms that deactivate or alert human operators when confidence in the updated model falls below a critical threshold [121]. As ML algorithms increasingly rely on cloud-based computing and distributed data pipelines, new points of vulnerability surface, such as data poisoning or adversarial inputs [122]. Consequently, interested researchers must explore cryptographic and blockchain-based solutions to secure the training process and validate inference stages [123]. These technologies may offer tamper-proof logs of training data usage and model updates and hence can offer a transparent chain of custody to strengthen a design's legal defensibility. Such strategies will likely gain traction in codes and standards as ML becomes more intertwined with operational decision-making.

In addition, liability risk management in ML-focused engineering contexts calls for interdisciplinary studies on insurance models [124]. Traditional professional liability insurance might not account for dynamic risk profiles introduced by continuous learning systems. Possible proposals could revolve around new insurance policies that specifically cover damage stemming from algorithmic miscalculations [125]. Collaboration between law and engineering researchers could also identify minimum performance guarantees that insurers would mandate before underwriting advanced ML-based projects.

## 7.0 Conclusions

Despite the potential and advantages of ML, navigating the legal and regulatory landscape requires engineers to reconcile this new technology with established liability doctrines, codes, and professional norms. This unique challenge demands more explicit standards and guidelines.

Liability doctrines such as negligence and product liability continue to serve as cornerstones for adjudicating disputes, yet they must evolve to account for the distributed risk profiles inherent in ML. Likewise, accountability grows more complex when third-party algorithmic developers and data suppliers assume partial roles in design workflows. Legislative bodies and professional societies are beginning to recognize these shifts, although explicit provisions for ML remain sparse in most building codes and regulations. In addition, this review points out the following:

- Performance-based codes offer a flexible avenue for introducing novel methods, but a broader consensus on development and validation procedures for ML models is necessary to ensure consistent levels of safety.
- Engineers, lawyers, policymakers, and data scientists are responsible for shaping a regulatory environment that fosters innovation and accountability.
- Future research directions highlight the need for standardized benchmarks, explainable frameworks, and multidisciplinary collaboration. These developments illuminate a larger theme: as technology evolves, so too must the ethical, legal, and professional scaffolding that underpins engineering practice.

**Data Availability**
Data is available on request from the author.

**Conflict of Interest**
The authors declare no conflict of interest.

**References**

[1]   K.M. Habibullah, J. Horkoff, Non-functional Requirements for Machine Learning: Understanding Current Use and Challenges in Industry, in: Proc. IEEE Int. Conf. Requir. Eng., 2021. https://doi.org/10.1109/RE51729.2021.00009.

[2]   P. Hacker, The European AI liability directives – Critique of a half-hearted approach and lessons for the future, Computer Law and Security Review. (2023). https://doi.org/10.1016/j.clsr.2023.105871.

[3]   Q. Chen, B. García de Soto, B.T. Adey, Construction automation: Research areas, industry concerns and suggestions for advancement, Automation in Construction. (2018). https://doi.org/10.1016/j.autcon.2018.05.028.

[4]   IBC, International Building Code, 2018. https://codes.iccsafe.org/content/IBC2018?site_type=public.

[5]   X. Xue, J. Zhang, Regulatory information transformation ruleset expansion to support automated building code compliance checking, Automation in Construction. (2022). https://doi.org/10.1016/j.autcon.2022.104230.

[6]   A.D. Selbst, Negligence and ai's human users, Boston University Law Review. (2020).

[7]   L. Al-Abdulkarim, K. Atkinson, T. Bench-Capon, A methodology for designing systems to reason with legal cases using Abstract Dialectical Frameworks, Artificial Intelligence and Law. (2016). https://doi.org/10.1007/s10506-016-9178-1.

[8]   K. Chagal-Feferkorn, How Can I Tell if My Algorithm Was Reasonable?, Michigan Technology Law Review. (2021). https://doi.org/10.36645/mtlr.27.2.how.

[9]   D.E. Bernstein, Frye, Frye, Again: The Past, Present, and Future of the General Acceptance Test, SSRN Electronic Journal. (2005). https://doi.org/10.2139/ssrn.262034.

[10]  P.B. Limpert, Beyond the Rule in Mohan: A New Model for Assessing the Reliability of Scientific Evidence, University of Toronto Faculty of Law Review. 54 (1996).

https://heinonline.org/HOL/Page?handle=hein.journals/utflr54&id=71&div=7&collection=journals (accessed February 9, 2025).

[11] A.W. Jurs, S. DeVito, Machines Like Me: A Proposal on the Admissibility of Artificially Intelligent Expert Testimony, (2024). https://papers.ssrn.com/abstract=4798356 (accessed February 9, 2025).

[12] H. Fraser, R. Simcock, A.J. Snoswell, AI Opacity and Explainability in Tort Litigation, in: ACM Int. Conf. Proceeding Ser., 2022. https://doi.org/10.1145/3531146.3533084.

[13] C.B. Robertson, Litigating Partial Autonomy, (2023). https://papers.ssrn.com/abstract=4392073 (accessed February 9, 2025).

[14] ASCE, AI and Civil Engineering | ASCE, (2025). https://www.asce.org/topics/ai-and-civil-engineering (accessed February 9, 2025).

[15] S. Engineering Institute of the American Society of Civil Engineers, A Vision for the Future of Structural Engineering and Structural Engineers: A Case for Change, 2019.

[16] IStructE, AI and the structural engineer - The Institution of Structural Engineers, (2025). https://www.istructe.org/resources/training/ai-and-the-structural-engineer/ (accessed February 9, 2025).

[17] I.D. Raji, P. Xu, C. Honigsberg, D. Ho, Outsider Oversight: Designing a Third Party Audit Ecosystem for AI Governance, in: AIES 2022 - Proc. 2022 AAAI/ACM Conf. AI, Ethics, Soc., 2022. https://doi.org/10.1145/3514094.3534181.

[18] A. Zapata, V.H. Menéndez, M.E. Prieto, C. Romero, A framework for recommendation in learning object repositories: An example of application in civil engineering, Advances in Engineering Software. (2013). https://doi.org/10.1016/j.advengsoft.2012.10.005.

[19] P.H.M. Sinsa, E.R. Savelsbergh, W.R. Van Joolingen, The difficult process of scientific modelling: An analysis of novices' reasoning during computer-based modelling, International Journal of Science Education. (2005). https://doi.org/10.1080/09500690500206408.

[20] C. Mennella, U. Maniscalco, G. De Pietro, M. Esposito, Ethical and regulatory challenges of AI technologies in healthcare: A narrative review, Heliyon. (2024). https://doi.org/10.1016/j.heliyon.2024.e26297.

[21] S. Mclachlan, M. Neil, K. Dube, R. Bogani, N. Fenton, B. Schaffer, Smart automotive technology adherence to the law: (de)constructing road rules for autonomous system development, verification and safety, International Journal of Law and Information Technology. (2021). https://doi.org/10.1093/ijlit/eaac002.

[22] E. Medina, Rethinking algorithmic regulation, Kybernetes. (2015). https://doi.org/10.1108/K-02-2015-0052.

[23] E. Karan, S. Asadi, Intelligent designer: A computational approach to automating design of windows in buildings, Automation in Construction. (2019). https://doi.org/10.1016/j.autcon.2019.02.019.

[24] P. Ohm, B. Reid, Regulating software when everything has software, in: George Washington Law Rev., 2016.

[25] B. Liu, Learning on the job: Online lifelong and continual learning, in: AAAI 2020 - 34th AAAI Conf. Artif. Intell., 2020. https://doi.org/10.1609/aaai.v34i09.7079.

[26] S. Singh, V. Rawat, D. Sharma, Tracing the Progression: Networking Applications From Conventional Methods to Incremental AI/ML Integration, Proceedings of International Conference on Communication, Computer Sciences and Engineering, IC3SE 2024. (2024) pp. 1753–1757. https://doi.org/10.1109/IC3SE62002.2024.10593172.

[27] A. Lavin, C.M. Gilligan-Lee, A. Visnjic, S. Ganju, D. Newman, S. Ganguly, D. Lange, A.G. Baydin, A. Sharma, A. Gibson, S. Zheng, E.P. Xing, C. Mattmann, J. Parr, Y. Gal, Technology readiness levels for machine learning systems, Nature Communications. (2022). https://doi.org/10.1038/s41467-022-33128-9.

[28]   J. Oviedo, M. Rodriguez, A. Trenta, D. Cannas, D. Natale, M. Piattini, ISO/IEC quality standards for AI engineering, Computer Science Review. 54 (2024) pp. 100681. https://doi.org/10.1016/J.COSREV.2024.100681.

[29]   N.N. Gutowski, NextGen Licensure & Accreditation, University of New Hampshire Law Review. 22 (2023). https://heinonline.org/HOL/Page?handle=hein.journals/plr22&id=325&div=16&collection=journals (accessed February 9, 2025).

[30]   I.H. Sarker, A.S.M. Kayes, S. Badsha, H. Alqahtani, P. Watters, A. Ng, Cybersecurity data science: an overview from machine learning perspective, Journal of Big Data. (2020). https://doi.org/10.1186/s40537-020-00318-5.

[31]   T. Alves, R. Das, T. Morris, Embedding Encryption and Machine Learning Intrusion Prevention Systems on Programmable Logic Controllers, IEEE Embedded Systems Letters. (2018). https://doi.org/10.1109/LES.2018.2823906.

[32]   G. Gharibi, V. Walunj, R. Nekadi, R. Marri, Y. Lee, Automated end-to-end management of the modeling lifecycle in deep learning, Empirical Software Engineering. (2021). https://doi.org/10.1007/s10664-020-09894-9.

[33]   S.K. Baduge, S. Thilakarathna, J.S. Perera, M. Arashpour, P. Sharafi, B. Teodosio, A. Shringi, P. Mendis, Artificial intelligence and smart vision for building and construction 4.0: Machine and deep learning methods and applications, Automation in Construction. (2022). https://doi.org/10.1016/j.autcon.2022.104440.

[34]   E. Vaughan, The Value and Impact of Building Codes, (n.d.).

[35]   S. Sattar, Evaluating the consistency between prescriptive and performance-based seismic design approaches for reinforced concrete moment frame buildings, Engineering Structures. (2018). https://doi.org/10.1016/j.engstruct.2018.07.080.

[36]   A. Paleyes, R.G. Urma, N.D. Lawrence, Challenges in Deploying Machine Learning: A Survey of Case Studies, ACM Computing Surveys. 55 (2022). https://doi.org/10.1145/3533378.

[37]   D. Lehr, P. Ohm, Playing with the Data: What Legal Scholars Should Learn About Machine Learning, UC Davis Law Review. (2017).

[38]   J. Delgado, L. Oyedele, A. Ajayi, L. Akanbi, L. Owolabi, Robotics and automated systems in construction: Understanding industry-specific challenges for adoption, Journal of Building Engineering. (2019).

[39]   Y. Grize, W. Fischer, C. Lützelschwab, Machine learning applications in nonlife insurance, Applied Stochastic Models in Business and Industry. (2020). https://doi.org/10.1002/asmb.2543.

[40]   BSI, European Committee for Standardization, Design of concrete structures - Part 1-2: General rules - Structural fire design, 2004. https://doi.org/10.1002/jcp.25002.

[41]   A. Alvarez, B.J. Meacham, N.A. Dembsey, J.R. Thomas, Twenty years of performance-based fire protection design: Challenges faced and a look ahead, Journal of Fire Protection Engineering. (2013). https://doi.org/10.1177/1042391513484911.

[42]   F. Tambon, G. Laberge, L. An, A. Nikanjam, P.S.N. Mindom, Y. Pequignot, F. Khomh, G. Antoniol, E. Merlo, F. Laviolette, How to certify machine learning based safety-critical systems? A systematic literature review, Automated Software Engineering. (2022). https://doi.org/10.1007/s10515-022-00337-x.

[43]   S.L. Brunton, J.N. Kutz, K. Manohar, A.Y. Aravkin, K. Morgansen, J. Klemisch, N. Goebel, J. Buttrick, J. Poskin, A.W. Blom-Schieber, T. Hogan, D. McDonald, Data-driven aerospace engineering: Reframing the industry with machine learning, AIAA Journal. (2021). https://doi.org/10.2514/1.J060131.

[44]   D.W. Johnston, N. Thomas Ahluwalia, M.B. Gwyn, Improving the Professional Engineering Licensure Process for Construction Engineers, Journal of Construction Engineering and Management. (2007). https://doi.org/10.1061/(asce)0733-9364(2007)133:9(669).

[45]   C. Musselman, J. Nelson, M. Phillips, Engineering Licensure Laws and Rules, Today and Tomorrow, in: 2020. https://doi.org/10.18260/1-2--17879.

[46]   A.A. Hopgood, Intelligent Systems for Engineers and Scientists : A Practical Guide to Artificial Intelligence, 2021.

[47]   V. Asghari, M. Hossein Kazemi, M. Shahrokhishahraki, P. Tang, A. Alvanchi, S.C. Hsu, Process-oriented guidelines for systematic improvement of supervised learning research in construction engineering, Advanced Engineering Informatics. (2023). https://doi.org/10.1016/j.aei.2023.102215.

[48]   B. Green, Y. Chen, The principles and limits of algorithm-in-the-loop decision making, Proceedings of the ACM on Human-Computer Interaction. (2019). https://doi.org/10.1145/3359152.

[49]   J. Hinze, Construction safety, (1997).

[50]   H. Prakken, Legal Reasoning: Computational Models, in: Int. Encycl. Soc. Behav. Sci. Second Ed., 2015. https://doi.org/10.1016/B978-0-08-097086-8.86161-9.

[51]   M.G. Farrell, Daubert v. Merrell Dow Pharmaceuticals, Inc.: Epistemilogy and Legal Process, Cardozo Law Review. 15 (1993). https://heinonline.org/HOL/Page?handle=hein.journals/cdozo15&id=2209&div=81&collection=journals (accessed February 9, 2025).

[52]   M.A. Schwartz, Kumho Tire Co. v. Carmichael: The Supreme Court Follows up on the Daubert Test, Touro Law Review. 16 (1999). https://heinonline.org/HOL/Page?handle=hein.journals/touro16&id=307&div=20&collection=journals (accessed February 9, 2025).

[53]   R.D. Marlin, Evidence - Scientific Evidence: Standard of Review Raises Questions of Fit. General Electric Co. v. Joiner, University of Arkansas at Little Rock Law Review. 21 (1998). https://heinonline.org/HOL/Page?handle=hein.journals/ualr21&id=143&div=13&collection=journals (accessed February 10, 2025).

[54]   R.A. Ragazzo, The Power of a Federal Appellate Court to Direct Entry of Judgment as a Matter of Law: Reflections on Weisgram v. Marley Co., Journal of Appellate Practice and Process. 3 (2001). https://heinonline.org/HOL/Page?handle=hein.journals/jappp3&id=121&div=13&collection=journals (accessed February 10, 2025).

[55]   K. Rasheed, A. Qayyum, M. Ghaly, A. Al-Fuqaha, A. Razi, J. Qadir, Explainable, trustworthy, and ethical machine learning for healthcare: A survey, Computers in Biology and Medicine. (2022). https://doi.org/10.1016/j.compbiomed.2022.106043.

[56]   California Business and Professions Code § 6700 (2024) :: 2024 California Code :: U.S. Codes and Statutes :: U.S. Law :: Justia, (n.d.). https://law.justia.com/codes/california/code-bpc/division-3/chapter-7/article-1/section-6700/ (accessed February 9, 2025).

[57]   A.M. Khan, A. BinZiad, A. Al Subaii, T. Alqarni, M.Y. Jelassi, A. Najmi, Supervised Learning Predictive Models for Automated Fracturing Treatment Design: A Workflow Based on Algorithm Comparison and Multiphysics Model Validation, in: Soc. Pet. Eng. - SPE Int. Hydraul. Fract. Technol. Conf. Exhib. IHFT 2022, 2022. https://doi.org/10.2118/205310-MS.

[58]   Eprs, Rapporteurs, BRIEFING EU Legislation in Progress Proposal for a regulation of the European Parliament and of the Council laying down harmonised rules on artificial intelligence (artificial intelligence act) and amending certain Union legislative acts Committees responsible, (n.d.).

[59]   K.W. Simons, A Restatement (Third) of Intentional Torts, Arizona Law Review. 48 (2006). https://heinonline.org/HOL/Page?handle=hein.journals/arz48&id=1077&div=59&collection=journals (accessed February 9, 2025).

[60]   C. Estlinbaum, Restitution Revisited, South Texas Law Review. 57 (2015).

https://heinonline.org/HOL/Page?handle=hein.journals/stexlr57&id=229&div=15&collection=journals (accessed February 10, 2025).

[61]    UNITED STATES COURT OF APPEALS FOR THE NINTH CIRCUIT, Estate of Barabin v. AstenJohnson, Inc., 740 F.3d 457 | Casetext Search + Citator, (2014). https://casetext.com/case/estate-of-barabin-v-astenjohnson-inc-3 (accessed February 10, 2025).

[62]    N.C. United States Court of Appeals, Primiano v. Cook, 598 F.3d 558 | Casetext Search + Citator, 2010. https://casetext.com/case/primiano-v-cook (accessed February 10, 2025).

[63]    A. Jobin, M. Ienca, E. Vayena, The global landscape of AI ethics guidelines, Nature Machine Intelligence. (2019). https://doi.org/10.1038/s42256-019-0088-2.

[64]    P. Huberman, Tort Law, Corrective Justice and the Problem of Autonomous-Machine-Caused Harm, Canadian Journal of Law and Jurisprudence. (2021). https://doi.org/10.1017/cjlj.2020.3.

[65]    J. Luiz De Moura, F. Júnior, Algorithmic torts, Brazilian Journal of Law, Technology and Innovation. 2 (2024) pp. 210–224. https://doi.org/10.59224/BJLTI.V2I1.210-224.

[66]    P. Watkins, E. Daniels, S. Slayton, First in the Nation: Arizona's Regulatory Sandbox, Stanford Law & Policy Review. (2018).

[67]    F.X. Zhao, J.L. Liu, Y. Zhou, Sandbox edge-based algorithm for multifractal analysis of complex networks, Chaos, Solitons and Fractals. (2023). https://doi.org/10.1016/j.chaos.2023.113719.

[68]    D. DeMatteo, K. Heilbrun, S. Arnold, A. Thornewill, Problem-Solving Courts and the Criminal Justice System, 2019. https://doi.org/10.1093/med-psych/9780190844820.001.0001.

[69]    B. Patten, Professional negligence in construction, 2012. https://doi.org/10.4324/9780203860526.

[70]    V.P. Goldberg, A Reexamination of "Glanzer v. Shepard": Surveyors on the Tort-Contract Boundary, Theoretical Inquiries in Law. (2009). https://doi.org/10.2202/1565-3404.1057.

[71]    M. Tafazzoli, K. Shrestha, H. Dang, Investigating Barriers to the Application of Automation in the Construction Industry, in: Constr. Res. Congr. 2024, CRC 2024, 2024. https://doi.org/10.1061/9780784485262.096.

[72]    C.W.J. Sorenson, Disclosure under Federal Rule of Civil Procedure 26(a)--Much Ado about Nothing, Hastings Law Journal. 46 (1994). https://heinonline.org/HOL/Page?handle=hein.journals/hastlj46&id=721&div=&collection= (accessed February 9, 2025).

[73]    W.A. Sinacori, Bily v. Arthur Young & (and) Co.: An Unnecessary Return to Privity in Cases of Auditor Negligence, Hofstra Property Law Journal. 6 (1993). https://heinonline.org/HOL/Page?handle=hein.journals/hofplj6&id=247&div=9&collection=journals (accessed February 10, 2025).

[74]    W.W.. Brady, Accountants' Liability to Third Parties : The Ultramares Case Reaffirmed, The Accounting Review. (1983).

[75]    P. Hacker, R. Krestel, S. Grundmann, F. Naumann, Explainable AI under contract and tort law: legal incentives and technical challenges, Artificial Intelligence and Law. (2020). https://doi.org/10.1007/s10506-020-09260-6.

[76]    C.A. Constantinides, Professional Ethics Codes in Court: Redefining the Social Contract between the Public and the Professions, Georgia Law Review. 25 (1990). https://heinonline.org/HOL/Page?handle=hein.journals/geolr25&id=1343&div=39&collection=journals (accessed February 9, 2025).

[77]    S. Roaf, Thermal landscaping of buildings : Climate-proofing design, Activism in Architecture. (2018) pp. 145–154. https://doi.org/10.4324/9781315182858-14.

[78]  Supreme Court of Pennsylvania, Bilt-Rite v. the Architectural Studio, 866 A.2d 270 | Casetext Search + Citator, 2005. https://casetext.com/case/bilt-rite-v-the-architectural-studio (accessed February 10, 2025).

[79]  A. Sethi, T. Ghosh, J. Lai, J. Zhang, Automation in Construction Contract Analysis and Management, in: Constr. Res. Congr. 2024, CRC 2024, 2024. https://doi.org/10.1061/9780784485262.063.

[80]  A. Di Benedetto, East River S.S. Corp. v. Transamerica Delaval, Inc.: Tort Or Contract - Answers for the Admirals, Loyola Law Review. 32 (1986). https://heinonline.org/HOL/Page?handle=hein.journals/loyolr32&id=1022&div=56&collection=journals (accessed February 10, 2025).

[81]  R.J. Traynor, G. V Yuba, UC Hastings Scholarship Repository Greenman v. Yuba Power Products, Inc. Recommended Citation, 2 (1963) pp. 57. http://repository.uchastings.edu/traynor_opinions/438 (accessed February 10, 2025).

[82]  C Mitcham, National Society of Professional Engineers (NSPE)... - Google Scholar, in: Encycl. Sci. Technol. Ethics, 2005. https://scholar.google.com/scholar?q=National+Society+of+Professional+Engineers+(NSPE)+Code+of+Eth ics+&hl=en&as_sdt=0,41 (accessed February 10, 2025).

[83]  M.E. Diamantis, Vicarious Liability for AI, Indiana Law Journal. 99 (2023). https://heinonline.org/HOL/Page?handle=hein.journals/indana99&id=330&div=11&collection=journals (accessed February 10, 2025).

[84]  S. Lupton, Cornes and Lupton's design liability in the construction industry: Fifth edition, 2013. https://doi.org/10.1002/9781444361162.

[85]  A.J. McNamara, S.M.E. Sepasgozar, Intelligent contract adoption in the construction industry: Concept development, Automation in Construction. (2021). https://doi.org/10.1016/j.autcon.2020.103452.

[86]  M.P. Gergen, Negligent Misrepresentation as Contract, California Law Review. 101 (2013). https://heinonline.org/HOL/Page?handle=hein.journals/calr101&id=995&div=26&collection=journals (accessed February 10, 2025).

[87]  K.S. Jordan, In re Caremark International, Inc. - Derivative Litigation, Del. Chancery, C.A. 13670 (Chancellor Allen, Sept. 15, 1996), Preventive Law Reporter. 15 (1996). https://heinonline.org/HOL/Page?handle=hein.journals/prevlr15&id=182&div=&collection= (accessed February 9, 2025).

[88]  C.A. Hill, B.H. McDonnell, Stone v. Ritter and the Expanding Duty of Loyalty, Fordham Law Review. 76 (2007). https://heinonline.org/HOL/Page?handle=hein.journals/flr76&id=1785&div=58&collection=journals (accessed February 10, 2025).

[89]  E. Ntoutsi, P. Fafalios, U. Gadiraju, V. Iosifidis, W. Nejdl, M.E. Vidal, S. Ruggieri, F. Turini, S. Papadopoulos, E. Krasanakis, I. Kompatsiaris, K. Kinder-Kurlanda, C. Wagner, F. Karimi, M. Fernandez, H. Alani, B. Berendt, T. Kruegel, C. Heinze, K. Broelemann, G. Kasneci, T. Tiropanis, S. Staab, Bias in data-driven artificial intelligence systems—An introductory survey, Wiley Interdisciplinary Reviews: Data Mining and Knowledge Discovery. (2020). https://doi.org/10.1002/widm.1356.

[90]  D.G. Johnson, Rethinking the Social Responsibilities of Engineers as a Form of Accountability, in: Philos. Eng. Technol., 2017. https://doi.org/10.1007/978-3-319-45193-0_7.

[91]  T. DoCarmo, S. Rea, E. Conaway, J. Emery, N. Raval, The law in computation: What machine learning, artificial intelligence, and big data mean for law and society scholarship, Law and Policy. (2021). https://doi.org/10.1111/lapo.12164.

[92]  B. Szabó, I. Babuška, Finite Element Analysis: Method, Verification and Validation, Second Edition, 2021. https://doi.org/10.1002/9781119426479.

[93]    H. Ben Braiek, F. Khomh, On testing machine learning programs, Journal of Systems and Software. (2020). https://doi.org/10.1016/j.jss.2020.110542.

[94]    Y.K. Wen, Reliability and performance-based design, Structural Safety. (2001). https://doi.org/10.1016/S0167-4730(02)00011-5.

[95]    G. Dulac-Arnold, N. Levine, D.J. Mankowitz, J. Li, C. Paduraru, S. Gowal, T. Hester, Challenges of real-world reinforcement learning: definitions, benchmarks and analysis, Machine Learning. (2021). https://doi.org/10.1007/s10994-021-05961-4.

[96]    J. Bandy, Problematic Machine Behavior: A Systematic Literature Review of Algorithm Audits, Proceedings of the ACM on Human-Computer Interaction. (2021). https://doi.org/10.1145/3449148.

[97]    K.L. Boyd, Datasheets for Datasets help ML Engineers Notice and Understand Ethical Issues in Training Data, Proceedings of the ACM on Human-Computer Interaction. (2021). https://doi.org/10.1145/3479582.

[98]    M.L. Rustad, T.H. Koenig, Towards a Global Data Privacy Standard, Florida Law Review. 71 (2019). https://heinonline.org/HOL/Page?handle=hein.journals/uflr71&id=381&div=14&collection=journals (accessed February 10, 2025).

[99]    C. Malhotra, V. Kotwal, S. Dalai, Ethical framework for machine learning, in: 10th ITU Acad. Conf. Kaleidosc. Mach. Learn. a 5G Futur. ITU K 2018, 2018. https://doi.org/10.23919/ITU-WT.2018.8597767.

[100]   D.R. Desai, J.A. Kroll, Trust but Verify: A Guide to Algorithms and the Law, Harvard Journal of Law & Technology (Harvard JOLT). 31 (2017). https://heinonline.org/HOL/Page?handle=hein.journals/hjlt31&id=7&div=4&collection=journals (accessed February 10, 2025).

[101]   E. Bardach, R.A. Kagan, Going by the book: The problem of regulatory unreasonableness, 2017. https://doi.org/10.4324/9780203790557.

[102]   I.D. Raji, A. Smart, R.N. White, M. Mitchell, T. Gebru, B. Hutchinson, J. Smith-Loud, D. Theron, P. Barnes, Closing the AI accountability gap: Defining an end-to-end framework for internal algorithmic auditing, in: FAT* 2020 - Proc. 2020 Conf. Fairness, Accountability, Transpar., 2020. https://doi.org/10.1145/3351095.3372873.

[103]   S. Li, X. Wang, A.S. Barnard, Diverse Explanations From Data-Driven and Domain-Driven Perspectives in the Physical Sciences, Machine Learning: Science and Technology. (2024). https://doi.org/10.1088/2632-2153/AD9137.

[104]   S. Burton, I. Habli, T. Lawton, J. McDermid, P. Morgan, Z. Porter, Mind the gaps: Assuring the safety of autonomous systems from an engineering, ethical, and legal perspective, Artificial Intelligence. (2020). https://doi.org/10.1016/j.artint.2019.103201.

[105]   H. Surden, Machine learning and law: An overview, in: Res. Handb. Big Data Law, 2021. https://doi.org/10.4337/9781788972826.00014.

[106]   A. Zhang, L. Xing, J. Zou, J.C. Wu, Shifting machine learning for healthcare from development to deployment and from models to data, Nature Biomedical Engineering. (2022). https://doi.org/10.1038/s41551-022-00898-y.

[107]   M. Anisetti, C.A. Ardagna, N. Bena, Continuous Certification of Non-functional Properties Across System Changes, in: Lect. Notes Comput. Sci. (Including Subser. Lect. Notes Artif. Intell. Lect. Notes Bioinformatics), 2023. https://doi.org/10.1007/978-3-031-48421-6_1.

[108]   A. Bertolini, F. Episcopo, The Expert Group's report on liability for artificial intelligence and other emerging digital technologies: A critical assessment, European Journal of Risk Regulation. (2021). https://doi.org/10.1017/err.2021.30.

[109]   R.K. Mothilal, A. Sharma, C. Tan, Explaining machine learning classifiers through diverse counterfactual explanations, in: FAT* 2020 - Proc. 2020 Conf. Fairness, Accountability, Transpar., 2020.

https://doi.org/10.1145/3351095.3372850.

[110] L. Tiberius Wiehler, How can AI regulation be effectively enforced? : comparing compliance mechanisms for AI regulation with a multiple-criteria decision analysis, (2022). https://doi.org/10.2870/5283385.

[111] Y. Zhou, Y. Yu, B. Ding, Towards MLOps: A Case Study of ML Pipeline Platform, in: Proc. - 2020 Int. Conf. Artif. Intell. Comput. Eng. ICAICE 2020, 2020. https://doi.org/10.1109/ICAICE51518.2020.00102.

[112] Y. Liu, M. Li, B.C.L. Wong, C.M. Chan, J.C.P. Cheng, V.J.L. Gan, BIM-BVBS integration with openBIM standards for automatic prefabrication of steel reinforcement, Automation in Construction. (2021). https://doi.org/10.1016/j.autcon.2021.103654.

[113] G. Al-Bdour, R. Al-Qurran, M. Al-Ayyoub, A. Shatnawi, Benchmarking open source deep learning frameworks, International Journal of Electrical and Computer Engineering. (2020). https://doi.org/10.11591/IJECE.V10I5.PP5479-5486.

[114] K.R. Varshney, H. Alemzadeh, On the Safety of Machine Learning: Cyber-Physical Systems, Decision Sciences, and Data Products, Big Data. (2017). https://doi.org/10.1089/big.2016.0051.

[115] W. Rodgers, R. Attah-Boakye, K. Adams, Application of Algorithmic Cognitive Decision Trust Modeling for Cyber Security Within Organisations, IEEE Transactions on Engineering Management. (2022). https://doi.org/10.1109/TEM.2020.3019218.

[116] E. Volokh, Chief Justice Robots, Duke Law Journal. (2019).

[117] R. Dwivedi, D. Dave, H. Naik, S. Singhal, R. Omer, P. Patel, B. Qian, Z. Wen, T. Shah, G. Morgan, R. Ranjan, Explainable AI (XAI): Core Ideas, Techniques, and Solutions, ACM Computing Surveys. (2023). https://doi.org/10.1145/3561048.

[118] G.E. Karniadakis, I.G. Kevrekidis, L. Lu, P. Perdikaris, S. Wang, L. Yang, Physics-informed machine learning, Nature Reviews Physics. (2021). https://doi.org/10.1038/s42254-021-00314-5.

[119] A. Nacef, A. Kaci, Y. Aklouf, D.L.C. Dutra, Machine learning based fast self optimized and life cycle management network, Computer Networks. (2022). https://doi.org/10.1016/j.comnet.2022.108895.

[120] S. De Vito, G. Di Francia, E. Esposito, S. Ferlito, F. Formisano, E. Massera, Adaptive machine learning strategies for network calibration of IoT smart air quality monitoring devices, Pattern Recognition Letters. (2020). https://doi.org/10.1016/j.patrec.2020.04.032.

[121] V. Campos, O. Klyagina, J.R. Andrade, R.J. Bessa, C. Gouveia, ML-assistant for human operators using alarm data to solve and classify faults in electrical grids, Electric Power Systems Research. 236 (2024) pp. 110886. https://doi.org/10.1016/j.EPSR.2024.110886.

[122] N. Papernot, P. McDaniel, A. Swami, R. Harang, Crafting adversarial input sequences for recurrent neural networks, in: Proc. - IEEE Mil. Commun. Conf. MILCOM, 2016. https://doi.org/10.1109/MILCOM.2016.7795300.

[123] M. Shen, X. Tang, L. Zhu, X. Du, M. Guizani, Privacy-Preserving Support Vector Machine Training over Blockchain-Based Encrypted IoT Data in Smart Cities, IEEE Internet of Things Journal. (2019). https://doi.org/10.1109/JIOT.2019.2901840.

[124] M. Baudry, C.Y. Robert, A machine learning approach for individual claims reserving in insurance, Applied Stochastic Models in Business and Industry. (2019). https://doi.org/10.1002/asmb.2455.

[125] N.G. Packin, Y. Lev-Aretz, Learning algorithms and discrimination, in: Res. Handb. Law Artif. Intell., 2018. https://doi.org/10.4337/9781786439055.00014.